\documentclass[twocolumn,aps,prd,10pt,showpacs,preprintnumbers,nofootinbib]{revtex4-1}
\usepackage{graphicx,slashed,bbold,subfigure,amsmath,amssymb,hyperref,enumerate}
\usepackage[utf8x]{inputenc}
\usepackage[english]{babel}

\begin{document}
\title{Strange quark matter in the presence of explicit symmetry breaking interactions} 
\author{J. Moreira}
\email{jmoreira@teor.fis.uc.pt}
\author{J. Morais}
\email{jorge.m.r.morais@gmail.com}
\author{B. Hiller}
\email{brigitte@teor.fis.uc.pt}
\author{A. A. Osipov}
\email{alexguest@teor.fis.uc.pt}
\author{A. H. Blin}
\email{alex@teor.fis.uc.pt}
\affiliation{Departamento de F\'{\i}sica da Universidade de Coimbra, 3004-516 Coimbra, Portugal}

\begin{abstract}
It is shown that a first order transition associated with a jump in the strange quark mass appears in a generalized 3 flavour Nambu--Jona-Lasinio (NJL) treatment of quark matter. The generalization of the Lagrangian displays the complete set of spin 0 interactions at leading and subleading orders  (LO and NLO) in $\frac{1}{N_c}$counting, including the recently derived  NLO explicit chiral symmetry breaking interactions which are of the same order as the 't Hooft flavour determinant. The parameters of the model are tightly constrained by the low energy characteristics in both the pseudoscalar and scalar meson sectors. The transition occurs in a moderate chemical potential region ($\mu \simeq 400~\mathrm{MeV}$ for zero temperature) in addition to the usual chiral transition associated with the light quark sector. This feature has at its root the inclusion of the explicit chiral symmetry breaking interactions, which therefore can be seen to act as a catalyst in the production of strange quark matter when compared to the conventional version of the model that takes only into account the 't Hooft interaction in the NLO. It can be traced back to the effect of the interactions which do not violate the Okubo-Zweig-Iizuka rule (OZI), without which the empirical ordering of the scalars ($m_{K^\star}< m_{a_0}\approx m_{f_0}$) is not reproduced.
\end{abstract}

%\begin{keyword}
%explicit chiral symmetry breaking \sep Nambu--Jona-Lasinio model \sep phase diagram \sep strange quark matter
%\end{keyword}

\maketitle

\section{Introduction}
In 1984 Witten \cite{Witten:1984rs} conjectures that stable and bound strange quark matter (SQM) may be formed under specific conditions (relevant early work in this idea can be found in \cite{Itoh:1970uw, Bodmer:1971we,Iachello:1974vx,Lee:1974ma,Terazawa:1979hqmod}). The validation of this hypothesis still remains a challenge as the region of baryonic chemical potential where it is expected to occur is of difficult access in heavy ion experiments and on lattice QCD calculations. The possibility of windows for stable SQM has been the subject of a significant number of theoretical approaches.

The properties of quark matter with comparable densities of up, down and strange quarks in beta equilibrium are studied in \cite{Farhi:1984qu} using the Fermi gas model and the MIT Bag model \cite{DeGrand:1975cf, Cleymans:1985wb} with extensions to include surface tension and Coulomb effects.

The subject of quark pair formation at finite densities, see \cite{Bailin:1983bm} for a review, was revived in more recent years with the understanding that at asymptotic high densities the ground state of cold matter is an electronless state necessarily composed of Cooper pairs built of quarks from all colours and flavours, the colour flavour locked (CFL) phase \cite{Alford:1998mk,Rapp:1999qa,Rajagopal:2000wf}. Away from this perturbative QCD regime a complex pattern of phases may occur, for reviews see \cite{Buballa:2003qv,Stephanov:2004xs,Schafer:2005ff,Schmitt:2010pn,Fukushima:2010bq}.

In \cite{Alford:2001zr} a model for the interface between the electron rich nuclear matter regime and the CFL phase is constructed. The windows in Bag model parameter space for SQM are shown to be much larger with the CFL pairing than without pairing \cite{Lugones:2002va}. 

While bound properties of SQM can be very conveniently addressed within Bag model studies, the aspects of SQM formation can be naturally addressed starting from a detailed scenario of chiral symmetry ($\chi_S$) breaking.

The seminal papers of Nambu--Jona-Lasinio \cite{Nambu:1961tp,Nambu:1961fr} present a model for dynamical chiral symmetry breaking with the fermion anti-fermion condensate as order parameter in the chiral limit (see also related works \cite{Nambu:1960tm,Nambu:1960xd,Vaks:1961,Arbuzov:1961}).
The model, although not renormalizable, and thus dependent on a cutoff, regained interest as an effective theory when 't Hooft derived a fermion interaction reminiscent of the NJL interaction, by eliminating the gluonic degrees of freedom of QCD in the semiclassical instanton approximation, which provides a solution to the $U_A(1)$ problem \cite{'tHooft:1976fv, PhysRevD.18.2199.3}. Furthermore he showd that QCD can be reduced to an effective meson theory in the large number of colours ($N_c$) limit \cite{'tHooft:1973jz}. 
Early derivations of effective meson interactions from the NJL model were obtained by its partial bosonization in \cite{Eguchi:1976iz,Kikkawa:1976fe}.
Physical masses for the related pseudoscalar Goldstones can be obtained through the inclusion, introduced in \cite{Volkov:1982zx}, of explicit $\chi_S$ breaking by QCD current quark mass terms.

Effective Lagrangians of low energy QCD operate at the scale of spontaneous chiral symmetry breaking, of the order of $\Lambda_{\chi SB}\sim 4\pi f_{\pi}$ \cite{Manohar:1983md}. In the NJL model  this scale is also related to the ultra-violet cutoff $\Lambda$ of the one-loop quark integral describing the gap equation, above which non-perturbative phenomena are expected to be less important. First results that relate the value of $\Lambda$ which consistently describes low-energy theorems and meson  observables to $\Lambda_{\chi SB}$ were obtained in \cite{Volkov:1982zx, Ebert:1982pk,Volkov:1984kq,Hatsuda:1984jm,Hatsuda:1985ey}.
Since then the model and its extensions are a landmark in the study of non-perturbative QCD effects (for reviews see for instance \cite{Klevansky:1992qe, Hatsuda:1994pi,Volkov:2006vq}).

Realistic NJL based investigations of SQM require the use of known extensions of the original Lagrangian, applied to three flavours, which include the $1/N_c$ suppressed NLO terms with multi-quark interactions. If one is interested only in the $U(1)_A$ breaking effects the extension includes the six-quark 't Hooft interactions (NJLH) \cite{'tHooft:1976fv, PhysRevD.18.2199.3, Kobayashi:1970ji,Kobayashi:1971qz}, considered first in the context of the NJL model in \cite{Kunihiro:1987bb,Bernard:1987sg, Reinhardt:1988xu}. In this form the model has been extensively applied (see for instance \cite{Kunihiro:1989my, Klimt:1989pm, Vogl:1989ea, Takizawa:1990ku, Vogl:1991qt,  Klevansky:1992qe, Bernard:1993wf,Hatsuda:1994pi, Birse:1996dx, Dmitrasinovic:2000ei, Naito:2003zj}).

A more detailed approach can also include  an appropriate set of $SU(3)_L\times SU(3)_R$ symmetric eight-quark interactions (NJLH8) \cite{Osipov:2005tq,Osipov:2006ns} completing the number of vertices  which are important in four dimensions for dynamical $\chi_S$ breaking as shown in \cite{Andrianov:1993wn,Andrianov:1992tb} in the mean field approximation. Furthermore the inclusion of these terms provides a solution to the instability of the NJLH effective potential \cite{Osipov:2005sp}. In the last decade the eight-quark extension of the model has received much attention (also in $SU(2)$ flavour space). Although the vacuum properties remain almost insensitive to the 8q interactions \cite{Osipov:2006ns}, their effects are remarkable for medium and thermal properties, see e.g. \cite{Osipov:2006ev,Osipov:2007mk,Kashiwa:2006rc,Kashiwa:2007hw,Hiller:2008nu,Sakai:2008py,Sakai:2010rp,Sasaki:2010jz,Bhattacharyya:2010wp,Bhattacharyya:2010ef,Bhattacharyya:2010jd}, as well as in the presence of a strong constant magnetic field  \cite{Osipov:2007je,Gatto:2010qs,Gatto:2010pt,Frasca:2011zn}. These findings point to the relevance of including higher dimension multi-quark interactions in effective models of the QCD ground state subject to external conditions. 

If one wants also to include at this order the effects of explicit $\chi_S$ symmetry breaking one should take advantage of the model recently presented in \cite{Osipov:2012kk},\cite{Osipov:2013fka}, the NJLH8m model, where m indicates that the NLO terms in current quark masses are taken into account in the effective multi-quark Lagrangian. 

Note that the role played by the current quark masses on phase transitions in hot and dense hadronic matter is presently a subject of intense study, as they are known to change the order of the transition or transform it into a smooth crossover \cite{Buballa:2003qv}.

A thorough understanding of the phase diagram in terms of the multi-quark interactions will benefit from further investigations in terms of a renormalization group approach, which lies, however, beyond the scope of the present paper.

In this work we  address for the first time the problem of the formation of a SQM phase using a detailed picture of the $\chi_S$ breaking assembled in the NJLH8m Lagrangian, in the mean field approximation. We review the model Lagrangian and introduce the thermodynamic potential at finite temperature, $T$, and baryonic chemical potential, $\mu$, in Section \ref{Model} and study the chiral $T-\mu$ phase diagram in Subsection \ref{PhaseDiagram}. The parameters of the model have been previously fixed at $T=\mu =0$ in such a way that the NJLH8m Lagrangian accurately describes the low-lying spectrum of the pseudoscalar and scalar mesons \cite{Osipov:2012kk}. They additionally provide a good description of the radiative two photon decays of the pseudoscalars and of the strong two body decays of the scalars \cite{Osipov:2013fka}. The success in the description of such a wide range of low energy hadronic properties could only be achieved through this apparently quite natural extension (NJLH8$\to$NJLH8m), which as we will see, has surprising consequences. In Subsection \ref{BetaEq} we address the case study for cold dense matter in $\beta$-equilibrium. The conclusions and further discussion of the obtained results are presented in Section \ref{Conclusions}.

\section{The Model}
\label{Model}
\subsection{ The effective Lagrangian }
\label{EffectL}

In this Section we give an overview of the Lagrangian which has been recently obtained in \cite{Osipov:2012kk,Osipov:2013fka}. Relying on the very general assumption that the scale $\Lambda$ determines the hierarchy of local multi-quark interactions which model QCD at low energy, we consider generic non-derivative vertices $L_i$ that contribute to the effective potential when $\Lambda\to\infty$,
\begin{align}
\label{genL}
   L_i\sim \frac{\bar g_i}{\Lambda^\gamma}\chi^\alpha\Sigma^\beta,
\end{align} 
where powers of $\Lambda$ give the correct dimensionality of the interactions. The couplings $\bar g_i$ are dimensionless whereas unbarred quantities $g_i=\frac{\bar g_i}{\Lambda^\gamma}$ will be used to denote the dimensionful couplings. The $L_i$ are C, P, T and chiral $SU(3)_L\times SU(3)_R$ invariant blocks, built of powers of the external sources $\chi$ and  of the  $U(3)$ Lie-algebra valued fields $\Sigma =(s_a-\imath p_a)\frac{1}{2}\lambda_a$, expressed in terms of the spin zero quark bilinears $s_a=\bar q\lambda_a q$, $p_a=\bar q\lambda_a \imath\gamma_5q$, and $a=0,1,\ldots ,8$, $\lambda_0=\sqrt{2/3}\times \mathbb{1}$, $\lambda_a$ being the standard $SU(3)$ Gell-Mann matrices for $1\leq a \leq 8$. With the sources $\chi$ having the same transformation properties as $\Sigma$, all the building blocks are in conformity with the symmetry constraints. At the end one is free to chose the external source to obtain a consistent set of explicit breaking chiral symmetry terms.

The interaction Lagrangian without external sources $\chi$ is well known, 
\begin{align}
\label{L-int}
L_{\mathrm{int}}&=\frac{\bar G}{\Lambda^2}\mbox{tr}\left(\Sigma^\dagger\Sigma\right)+\frac{\bar\kappa}{\Lambda^5}\left(\det\Sigma+\det\Sigma^\dagger\right) \nonumber \\
&+\frac{\bar g_1}{\Lambda^8}\left(\mbox{tr}\,\Sigma^\dagger\Sigma\right)^2+\frac{\bar g_2}{\Lambda^8}\mbox{tr}\left(\Sigma^\dagger\Sigma\Sigma^\dagger\Sigma\right).
\end{align}  
It contains four dimensionful couplings $G, \kappa, g_1, g_2$, which refer to the NJL, 't Hooft flavour determinant and eight-quark interactions referred to previously. 

Turning now to the source dependent vertices, the following eleven terms were obtained in  \cite{Osipov:2012kk,Osipov:2013fka} for the generalized effective multi-quark Lagrangian, 
\begin{align}
   L_\chi &=\sum_{i=0}^{10}L_i, \nonumber \\
   L_0&=-\mbox{tr}\left(\Sigma^\dagger\chi +\chi^\dagger\Sigma\right), \nonumber \\
   L_1&=-\frac{\bar\kappa_1}{\Lambda}e_{ijk}e_{mnl}\Sigma_{im}\chi_{jn}\chi_{kl}+h.c.\nonumber \\
   L_2&=\frac{\bar\kappa_2}{\Lambda^3}e_{ijk}e_{mnl}\chi_{im}\Sigma_{jn}\Sigma_{kl}+h.c.,\nonumber \\
   L_3&=\frac{\bar g_3}{\Lambda^6}\mbox{tr}\left(\Sigma^\dagger\Sigma\Sigma^\dagger\chi\right)+h.c.\nonumber \\
   L_4&=\frac{\bar g_4}{\Lambda^6}\mbox{tr}\left(\Sigma^\dagger\Sigma\right)\mbox{tr}\left(\Sigma^\dagger\chi\right)+h.c.,\nonumber \\
   L_5&=\frac{\bar g_5}{\Lambda^4}\mbox{tr}\left(\Sigma^\dagger\chi\Sigma^\dagger\chi\right)+h.c.\nonumber \\
   L_6&=\frac{\bar g_6}{\Lambda^4}\mbox{tr}\left(\Sigma\Sigma^\dagger\chi\chi^\dagger +\Sigma^\dagger\Sigma\chi^\dagger\chi\right),\nonumber \\
   L_7&=\frac{\bar g_7}{\Lambda^4}\left(\mbox{tr}\Sigma^\dagger\chi+ h.c.\right)^2\nonumber \\ 
   L_8&=\frac{\bar g_8}{\Lambda^4}\left(\mbox{tr}\Sigma^\dagger\chi- h.c.\right)^2,\nonumber \\ 
   L_9&=-\frac{\bar g_9}{\Lambda^2}\mbox{tr}\left(\Sigma^\dagger\chi\chi^\dagger\chi\right)+h.c.\nonumber \\
   L_{10}&=-\frac{\bar g_{10}}{\Lambda^2}\mbox{tr}\left(\chi^\dagger\chi\right)\mbox{tr}\left(\chi^\dagger\Sigma\right)+h.c.
\end{align}
Putting finally $\chi 
=\frac{1}{2}\mbox{diag}(\nu_u, \nu_d, \nu_s)$ this set contains all the explicitly breaking chiral symmetry terms relevant at the scale of spontaneous symmetry breaking. 

The standard mass term of the free Lagrangian (contained here in $L_0$) is only a part of a more complicated picture arising in effective models beyond leading order \cite{Gasser:1982ap}. Chiral perturbation theory \cite{Weinberg:1978kz,Pagels:1974se,Gasser:1983yg,Gasser:1984gg} is a well-known example of a self consistent assembling of the mass terms, order by order, in an expansion in the masses themselves.

From the point of view of symmetry content, were the $\Sigma$ fields chosen to represent pointlike spin-0 mesonic fields (of energy dimension $\left[\Sigma\right]=E$) instead of the quark bilinears ($\left[\Sigma\right]=E^3$), $L_\mathrm{int}$ (with accordingly adjusted dimensions of the couplings) would correspond to the Linear Sigma Model \cite{GellMann:1960np} interaction Lagrangian without fermions and the addition of $L_\chi$ would provide a generalization thereof. The composite/structureless nature of the fields has however deep consequences for the difference in dynamics and renormalization of both approaches.

We thus get the effective multi-quark Lagrangian relevant at the scale of dynamical chiral symmetry breaking as 
\begin{equation}
\label{Leff}
L_{\mathrm{eff}}=\imath{\bar q} \partial^{\mu} \gamma_{\mu} q + L_{\mathrm{int}}+L_{\chi}.
\end{equation}
One may ask how it relates to the powerful large $N_c$ classification. We have obtained (for details please see  \cite{Osipov:2012kk,Osipov:2013fka}) that exactly the diagrams which survive as $\Lambda\rightarrow\infty$ also surive as $N_c\rightarrow\infty$ with the following assignments: $\Sigma \sim N_c$; $\Lambda \sim N_c^0 \sim 1$; $\chi \sim N_c^0 \sim 1$ \footnote{The counting for $\Lambda$ is a direct consequence of the gap equation $1\sim N_c G\Lambda^2$.}.

The following usual requirements are fulfilled:
\begin{enumerate}[(i)]
\item the leading quark contribution to the vacuum energy from $4q$ interactions is of order $N_c$, thus $G\sim \frac{1}{N_c}$; 
\item the $6q$ $U_A(1)$ anomaly is suppressed by one power of $\frac{1}{N_c}$,  $\kappa \sim \frac{1}{N_c^3}$; 
\item Zweig's rule violating effects are always of order $\frac{1}{N_c}$ with respect to the leading contribution.
\item Non OZI-violating Lagrangian pieces scaling as $N_c^0$ represent NLO contributions with one internal quark loop in $N_c$ counting; their couplings encode the admixture of a four quark component ${\bar q}q{\bar q}q$ to the leading ${\bar q}q$ contribution at $N_c\rightarrow\infty$.
\end{enumerate}

At LO in $1/N_c$ only the  $4q$ interactions $(\sim G)$ in eq. (\ref{L-int}) and $L_0$ contribute. All other terms in the Lagrangian are NLO and of the same order as the 't Hooft determinant. As a consequence a whole plethora of interactions comes into play, competing with or enhancing the effects of the axial $U(1)_A$ symmetry breaking ascribed to the 't Hooft determinant. The NLO interactions are classified in two groups, the ones that trace the Zweig's rule violation and are proportional to $\kappa,\kappa_1,\kappa_2,g_1,g_4,g_7,g_8,g_{10}$ and those with the admixture of $\overline{q}q\overline{q}q$ and $\overline{q}q$ quark states going with $g_2,g_3,g_5,g_6,g_9$. 

These interactions are of the utmost importance for the successful fit of the scalar spectrum and allow us to establish a bridge between our Lagrangian approach and the  methods which consider explicitly meson loop corrections, tetraquark
configurations and so on \cite{Jaffe:1976ig, Jaffe:1976ih,Black:1998wt,Wong:1980je,Narison:1986vw,vanBeveren:1986ea, Latorre:1985uy,
%Narison:1996fm,
Achasov:1982bt,Weinstein:1990gu,
Alford:2000mm,
Fariborz:2008bd,Fariborz:2009cq,Close:2002zu,Klempt:2007cp}.

From the 18 model parameters, 3 of them ($\kappa_1, g_9, g_{10}$) contribute to the current quark masses $m_i, i=u,d,s$ and express the Kaplan-Manohar ambiguity \cite{Kaplan:1986ru}. They can be set to 0 without loss of generality leading to $\nu_i=m_i$. One ends up with 5 parameters needed to describe the LO contributions (the scale $\Lambda$, the coupling $G$, and the $m_i$)  and 10 in NLO ( $\kappa, \kappa_2$, $g_1,\ldots, g_{8}$). The increase in the number of parameters at NLO is a common feature in any effective Lagrangian approach. The parameters are controlled on the theoretical side through the symmetries of the Lagrangian and on the phenomenological side through the low energy characteristics of the pseudoscalar and the scalar mesons. 

The motivation for the present work resides in the desire to understand the impact of the new NJLH8m terms in the thermodynamical properties of the model. Once one accepts that the contribution of the subleading in $N_c$ counting  't Hooft interaction can be taken at tree mesonic level without consideration of mesonic loop corrections which act at the same subleading order in Nc (the usual approach),  it is not only reasonable but mandatory to  include at the same level of approximation all terms that contribute at the same order as the 't Hooft term  in the tree level mesonic Lagrangian. The terms which are taken into account are the leading ones in their class.

The bosonization of the quark Lagrangian (\ref{Leff}) is carried out by path integral techniques. Using a functional identity  \cite{Reinhardt:1988xu} a set of physical scalar and pseudoscalar fields $\sigma =\sigma_a
\lambda_a,\,\phi = \phi_a\lambda_a$ is introduced in addition to the auxiliary bosonic variables 
 $s_a=\bar q\lambda_aq,\, p_a=\bar qi\gamma_5
\lambda_aq$ and the Lagrangian is equivalently rewritten in a semi-bosonized form as 
%the following bilinear form in quark fields and a polynomial dependence on the bosonic varables
\begin{align}
\label{L}
   L&=\bar q\left(i\gamma^\mu\partial_\mu -\sigma - i\gamma_5\phi\right)q + L_{\mathrm{aux}},\\
   L_{\mathrm{aux}}&=s_a\sigma_a + p_a\phi_a - s_am_a +L_{\mathrm{int}}(s,p) +\sum_{i=2}^8L^\prime_i(s,p,m) \nonumber,
\end{align}
where the last sum collects all terms of $L_\chi$ expressed with auxiliary variables $s$, $p$, except $L_0$ (now explicitly shown as $s_a m_a$) and the terms which are set to zero using the freedom given by the Kaplan-Manohar ambiguity.

Integrating out the quarks and the $s_a,p_a$ fields and after performing a shift in the scalar field $\sigma\to \sigma +{\cal M}$, such that the new $\sigma$-field has a vanishing vacuum expectation value, $\langle\sigma\rangle =0$, in the spontaneously broken phase, one obtains an effective mesonic Lagrangian,
\begin{align}
\label{bos}
\mathcal{L}_{\mathrm{bos}}=&\mathcal{L}_{\mathrm{st}}+\mathcal{L}_{\mathrm{hk}}, \nonumber \\
\mathcal{L}_{\mathrm{st}}=&h_a\sigma_a+\frac{h_{ab}^{(1)}}{2} \sigma_a\sigma_b+\frac{h_{ab}^{(2)}}{2} \phi_a\phi_b  \nonumber \\
&+\sigma_a(\frac{1}{3}+h^{(1)}_{abc}\sigma_b\sigma_c +h^{(2)}_{abc}\phi_b\phi_c)+{\cal O}({\mbox{f\mbox{}ield}}^4),\nonumber\\
\int\!\mathrm{d}^4 x_E \mathcal{L}_{\mathrm{hk}}=&\frac{1}{2}\mbox{ln}|\mathrm{det} D^{\dagger}_E D_E|=W_{\mathrm{hk}}(\sigma,\phi )\nonumber \\
=&-\!\int\!\frac{\mathrm{d}^4 x_E}{32 \pi^2}\sum_{i=0}^\infty I_{i-1}\mbox{tr}(b_i) ,\nonumber\\
b_0 =&1,\hspace{0.2cm} b_1=-Y,\nonumber\\
b_2=&\frac{Y^2}{2} +\frac{\lambda_3 }{2}\Delta_{ud}Y+\frac{\lambda_8}{2\sqrt{3}} (\Delta_{us}+\Delta_{ds})Y, \hspace{0.2cm} \ldots , \nonumber\\
Y=&i\gamma_{\alpha}(\partial_{\alpha}\sigma +i\gamma_5\partial_{\alpha}\phi )+\sigma^2+\{{\cal M},\sigma\}+\phi^2 +\nonumber \\
&i\gamma_5[\sigma+{\cal M},\phi ],
\end{align}
with the constituent quark mass matrix denoted by ${\cal M}=\mbox{diag}(M_u,M_d,M_s)$ and $\Delta_{ij}=M_i^2-M_j^2$.

The $\mathcal{L}_\mathrm{st}$ is the result of the stationary phase integration at leading order over the auxiliary bosonic variables $s_a,p_a$, shown as a series in growing powers of $\sigma_a$ and $\phi_a$.  The coefficients $h^{(i)}_{ab...}$ in $\mathcal{L}_{st}$ are obtained recursively from $h_a$ (which are related to the condensates).

The quantity $W_\mathrm{hk}$ results from the Gaussian integration over the quark bilinear form in the heat kernel approach, that appropriately takes into account the quark mass differences \cite{Osipov:2000rg,Osipov:2001bj}. The Laplacian in euclidean space-time ${D^{\dagger}_E D}_E={\cal M}^2-\partial_\alpha^2+Y$ is associated with the euclidean Dirac operator $D_E=i\gamma_\alpha\partial_\alpha -{\cal M}-\sigma -i\gamma_5\phi$. The kinetic terms for the $\sigma$ and $\phi$ fields are generated radiatively through $W_{hk}$.
The quantities $I_i$  are the arithmetic averages $I_i=\frac{1}{3}\sum_{f=u,d,s}J_i(M_f^2)$ over the 1-loop euclidean momentum integrals $J_i$ with $i+1$ vertices ($i=0,1,\ldots$) :
\begin{align} 
\label{Ji}
    J_i(M^2)=&16\pi^2\Gamma (i+1)\!\int  \frac{\mathrm{d}^4p_E}{(2\pi)^4}\,\hat\rho_\Lambda \frac{1}{(p_E^2+M^2)^{i+1}}\\ 
		        =&16\pi^2\!\int  \frac{\mathrm{d}^4p_E}{(2\pi)^4} \int^\infty_0 \mathrm{d}\tau \,\hat\rho_\Lambda\tau^{i}\mathrm{e}^{-\tau (p_E^2+M^2)}. \nonumber
\end{align}  

They can be evaluated with a covariant regulator,
\begin{align}
\hat\rho_\Lambda\left(\tau\Lambda^2\right)=1-(1+\tau\Lambda^2)\mathrm{e}^{-\tau\Lambda^2},
\end{align}
corresponding to two Pauli-Villars subtractions in the integrand \cite{Pauli:1949zm,Osipov:1985} (previously used for instance in \cite{Osipov:2003xu,Osipov:2006ns,Osipov:2006xa, Osipov:2007mk,Osipov:2012kk, Osipov:2013fka}).

We consider only the dominant contributions to the heat kernel series, up to $b_2$ for the meson spectra and strong decays. These involve the quadratic and logarithmic in $\Lambda$ quark loop integrals $I_0$ and $I_1$ respectively. The $J_{-1}$ integral can be easily obtained through the relation:
\begin{align}
 J_{-1}(M^2)=-\int_0^{M^2}J_0(\alpha^2)\mathrm{d}\alpha^2.
\end{align}

\subsection{Model thermodynamic potential}
\label{ModelPotential}
The thermodynamic potential of the model Lagrangian is given in the mean field approximation by:
\begin{align}
\label{Omega}
\Omega=&\mathcal{V}_{st}+\sum_i\frac{N_c}{8\pi^2}J_{-1}(M_i,T,\mu_i),
\end{align}
where the index refers to the sum over flavours $i=u,d,s$. 

The Dirac and Fermi sea contributions to the part stemming from the fermionic path integral, $J^{vac}_{-1}$ and $J^{med}_{-1}$, can be written as \cite{Osipov:2003xu, Hiller:2008nu}:
\begin{align}
J_{-1}=&J^{vac}_{-1}+J^{med}_{-1},\nonumber\\
J^{vac}_{-1}=&\int\frac{\mathrm{d}^4 p_E}{(2\pi)^4}\int^\infty_0 \frac{\mathrm{d}\tau}{\tau}\rho\left(\tau\Lambda^2\right)16\pi^2\nonumber\\
&\times\left(\mathrm{e}^{-\tau\left(p_ {0\,E}^2+\boldsymbol{p}^2+M^2\right)}-\mathrm{e}^{-\tau\left(p_ {0\,E}^2+\boldsymbol{p}^2\right)}\right),\nonumber\\
J^{med}_{-1}=&-\int\frac{\mathrm{d}^3p}{(2\pi)^3}16\pi^2T\left.\left(\mathcal{Z}^++\mathcal{Z}^-\right)\right|^{M}_{0}+C(T,\mu),\nonumber\\
\mathcal{Z}^\pm =&\mathrm{log}\left(1+\mathrm{e}^{-\frac{E\mp\mu}{T}}\right)-\mathrm{log}\left(1+\mathrm{e}^{-\frac{E_\Lambda\mp\mu}{T}}\right)\nonumber\\
&-\frac{\Lambda^2}{2T E_\Lambda}\frac{\mathrm{e}^{-\frac{E_\Lambda\mp\mu}{T}}}{1+\mathrm{e}^{-\frac{E_\Lambda\mp\mu}{T}}},\nonumber\\
C(T,\mu)=&\int\frac{\mathrm{d}^3p}{(2\pi)^3}16\pi^2T~\nonumber\\
&\times\mathrm{log}\left(\left(1+\mathrm{e}^{-\frac{|\boldsymbol{p}|-\mu}{T}}\right)\left(1+\mathrm{e}^{-\frac{|\boldsymbol{p}|+\mu}{T}}\right)\right)
\end{align}
where $E=\sqrt{M^2+\boldsymbol{p}^2}$ and $E_\Lambda=\sqrt{E^2+\Lambda^2}$. The $|^{M}_{0}$ notation refers to the subtraction of the same quantity evaluated for $M=0$, which is done so as to set the zero of the potential to a uniform gas of massless quarks (it amounts to a subtraction of a constant). The $C(T,\mu)$ term is needed for thermodynamic consistency \cite{Hiller:2008nu}%,is set to counterbalance the first part of the $\mathcal{Z^\pm_\pm}$ terms in the zero mass subtractions

The stationary phase contribution coming from the integration over the auxiliary bosonic fields, $\mathcal{V}_{st}$, is evaluated using standard techniques \cite{Osipov:2003xu} and is given by:
\begin{align}
\label{Vst}
\mathcal{V}_{st}=&\nonumber\\
\frac{1}{16}
\bigg(&  4 G \left(h_i^2\right) + 3 g_1 \left(h_i^2\right)^2 + 3 g_2 \left(h_i^4\right) + 4 g_3 \left(h_i^3 m_i\right)\nonumber\\
			&+ 4 g_4 \left(h_i^2\right) \left(h_j m_j\right) + 2 g_5 \left(h_i^2 m_i^2\right) + 2 g_6 \left(h_i^2 m_i^2\right) \nonumber\\ 
	    &+ 4 g_7 \left(h_i m_i\right)^2 + 8 \kappa h_u h_d h_s \nonumber\\
			&+ 8 \kappa_2 \left( m_u h_d h_s + h_u m_d h_s +  h_u h_d m_s\right)\bigg)\bigg|^{M_i}_0,
\end{align}
where $h_i$ ($i=u,d,s$) are twice the quark condensates and $m_i$ the current masses (a summation over the $i,~j$ flavour indices in $\mathcal{V}_{st}$ is implicit). 

The stationary phase conditions which relate the dynamical masses to the condensates,
\begin{align}
\label{spc}
\Delta_f =&  M_f-m_f\\ 
         =& -G h_f -\frac{g_1}{2} h_f(h_i^2) - \frac{g_2}{2} (h_f^3) - \frac{3 g_3}{4} h_f^2 m_f\nonumber\\
				  &	-\frac{  g_4}{4} \left( m_f \left(h_i^2\right)+2 h_f(m_i h_i)\right) - \frac{g_5 + g_6}{2} h_f m_f^2\nonumber\\
					&	- g_7 m_f (h_i m_i) - \frac{\kappa}{4} t_{fij}h_i h_j - \kappa_2 t_{fij}h_i m_j,\nonumber
\end{align}
where $t_{fij}$ corresponds to the totally symmetric quantity with nonzero components $t_{uds}=1$, are solved self-consistently with the gap equations which correspond to the minimization of the thermodynamic potential (implicit summation over $i,~j$). 

\section{Results and discussion}
\label{Results}
\subsection{\texorpdfstring{$T-\mu$}{T-mu} phase diagram}
\label{PhaseDiagram}
\begin{figure*}
\center
\subfigure[]{\label{grafPDD2}\includegraphics[width=0.32\textwidth]{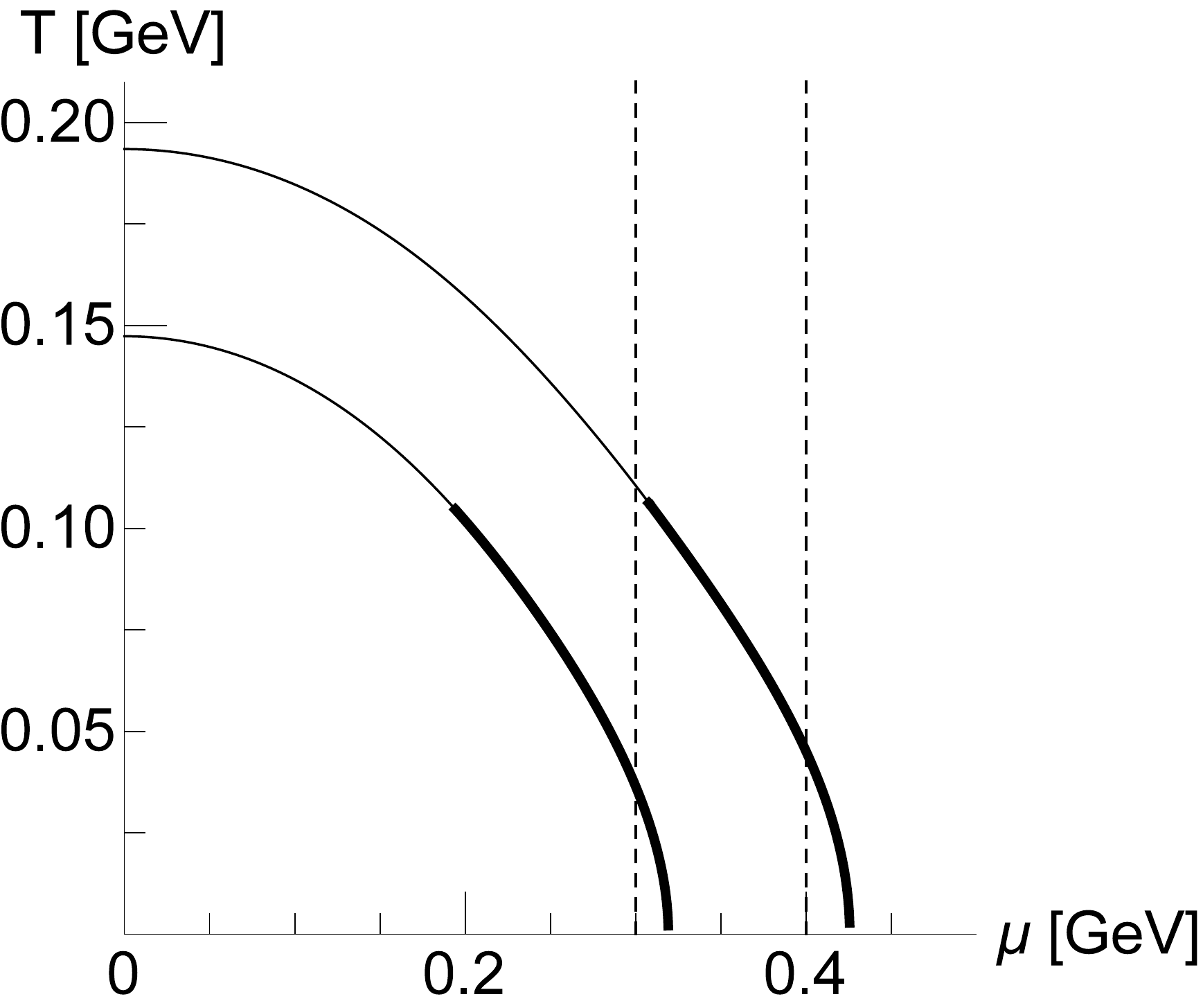}}
\subfigure[]{\label{grafMuMsmu0300}\includegraphics[width=0.32\textwidth]{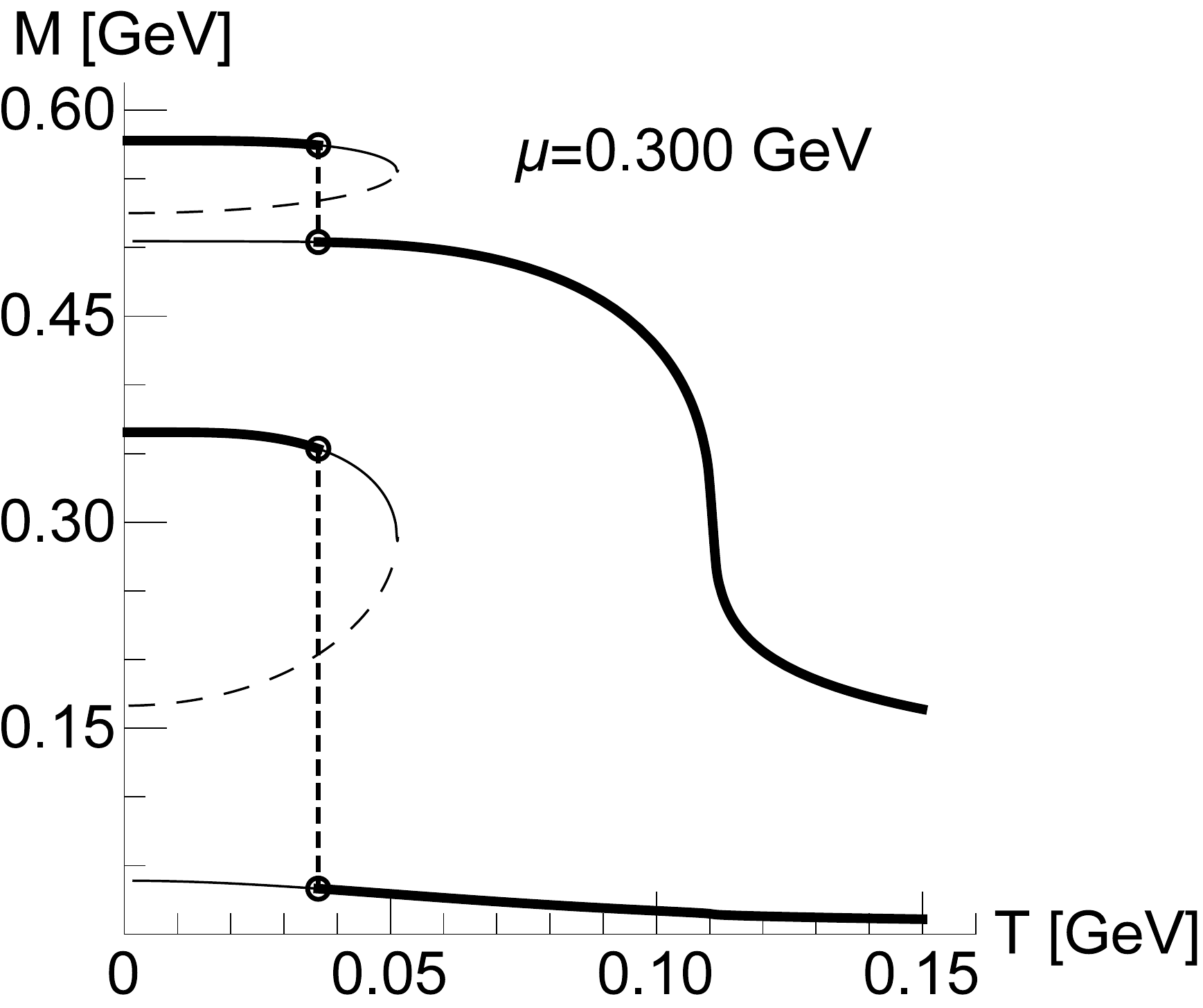}}
\subfigure[]{\label{grafMuMsmu0400}\includegraphics[width=0.32\textwidth]{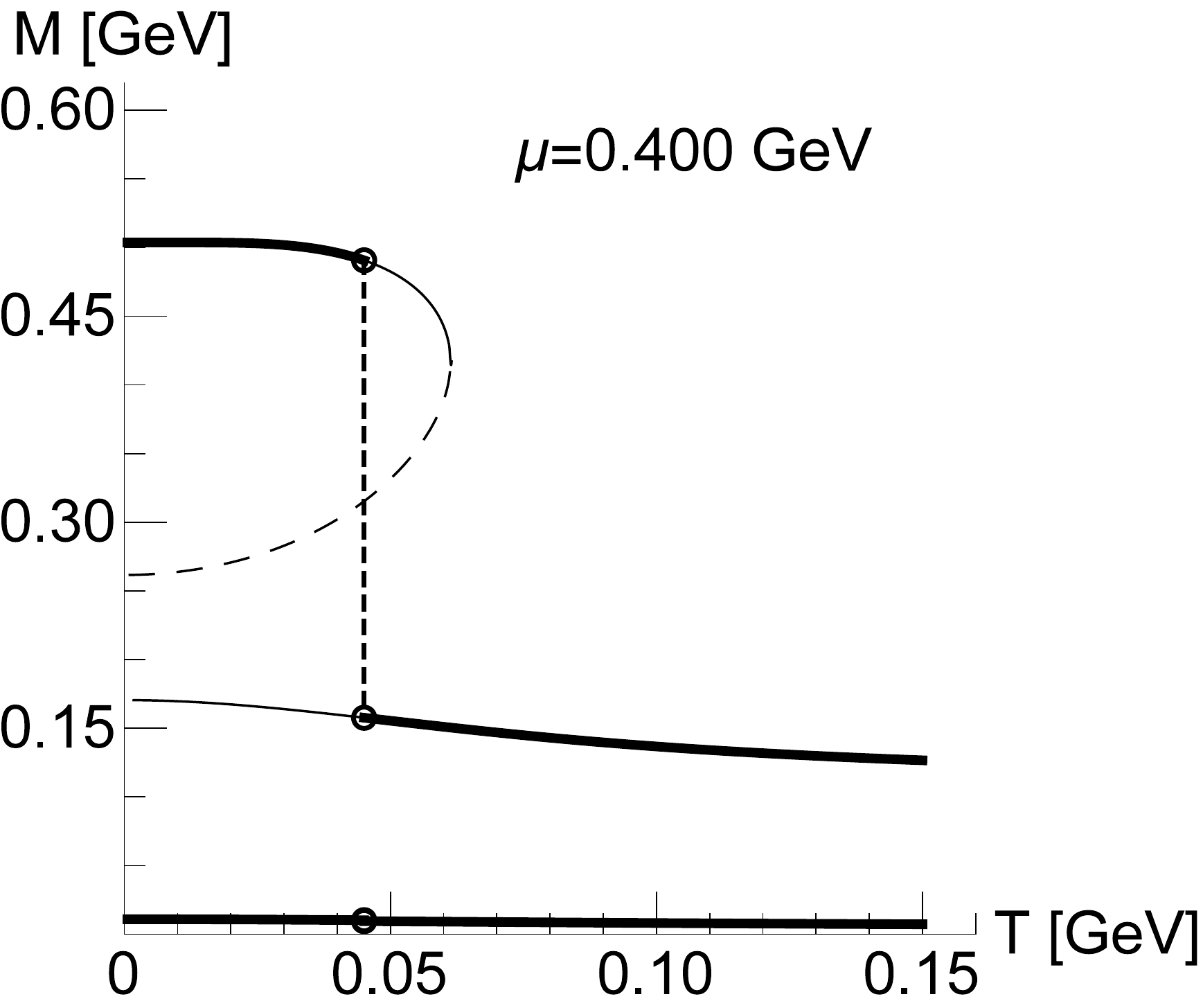}}
\caption{
Phase diagram in the $\mu-T$ plane using the parameter set from Table \ref{parametersetD2} displaying the two transition lines (first order/crossover corresponds to the thick/thin lines) associated with the light and strange quarks (outer refers to the strange) in \ref{grafPDD2}. Dynamical mass profiles as a function of temperature at the two chemical potentials indicated in \ref{grafPDD2} by the vertical lines: $\mu=0.300~\mathrm{GeV}$ in Fig. \ref{grafMuMsmu0300} and $\mu=0.400~\mathrm{GeV}$ in Fig. \ref{grafMuMsmu0400}. The upper two lines are for $M_s$, the lower ones for $M_l$; thick lines represent physical solutions, thin and dashed lines show relative minima and maxima of the thermodynamic potential, respectively;  first-order transitions are represented as dashed lines connecting two circles. In these figures, $\mu_u = \mu_d = \mu_s = \mu$.
}
\label{Imagens_PhaseDiagram}
\end{figure*}

Fitting the model parameters to properties of the low lying scalars and pseudoscalars (as seen in Table \ref{parametersetD2}) the most remarkable feature of the $\mu-T$ phase diagram, when compared to NJLH and NJLH8 (\cite{Osipov:2006ev, Hiller:2008nu,Osipov:2007mk}), is the appearance of a second 1st order transition line\footnote{In the NJLH model, two critical lines occur only for unphysically small values of $\kappa$ and $m_u\ne m_s$ \cite{Buballa:2003qv} otherwise a crossover behaviour for the 2nd transition prevails.} as seen in Fig.\ref{Imagens_PhaseDiagram}. A realistic fit to the spectra appears to imply this feature as it also occurs when using the parameter sets previously reported in \cite{Osipov:2013fka} (see Table \ref{CEPS}). Furthermore it is also unaffected by a change in the regularization scheme as is exemplified by the choice of a 3D cutoff which results in a relatively small shift of the first order transition lines to higher chemical potential (see parameter set f in Table \ref{CEPS}). For the remaining results presented in this paper the Pauli-Villars regularization scheme with the parameter set from the top row of Table \ref{parametersetD2} is used.The appearance of this additional line is somewhat surprising as usually an increase in finite current mass terms has the effect of  smoothing out the transition behaviour.

\begin{table*}
\caption{Model parameters obtained using a regularization kernel with two Pauli-Villars subtractions in the integrand and using a 3D cutoff, in the first and second row respectively (see \cite{Osipov:1985,Moreira:2010bx}), given in the following units: $\left[\Lambda\right]=\text{MeV}$, $\left[G\right]=\text{GeV}^{-2}$, $\left[\kappa_2\right]=\text{GeV}^{-3}$, $\left[g_5\right]=\left[g_6\right]=\left[g_7\right]=\left[g_8\right]=\text{GeV}^{-4}$,  $\left[\kappa\right]=\text{GeV}^{-5}$ and $\left[g_3\right]=\left[g_4\right]=\text{GeV}^{-6}$, $\left[g_1\right]=\left[g_2\right]=\text{GeV}^{-8}$. For their fitting we used the current quark masses values  $m_u=4$ MeV, $m_s=100$ MeV and the empirical input presented in the bottom row (meson masses and weak decays in MeV, pseudoscalar and scalar mixing angles, $\theta_{ps}$ and $\theta_{s}$, in degrees). From the self-consistent resolution of the gap equations we obtain $M_u=375~\mathrm{MeV}$, $M_s=546~\mathrm{MeV}$ for the constituent quark masses in the PV regularization scheme and $M_u=394~\mathrm{MeV}$, $M_s=637~\mathrm{MeV}$ in the 3D case.}
\label{parametersetD2}
\begin{tabular*}{\textwidth}{@{\extracolsep{\fill}}rrrrrrrrrrrr@{}}\hline
  \multicolumn{1}{c}{$G$}
& \multicolumn{1}{c}{$\kappa$}
& \multicolumn{1}{c}{$\kappa_2$}
& \multicolumn{1}{c}{$g_1$}
& \multicolumn{1}{c}{$g_2$}
& \multicolumn{1}{c}{$g_3$}
& \multicolumn{1}{c}{$g_4$}
& \multicolumn{1}{c}{$g_5$}
& \multicolumn{1}{c}{$g_6$}  
& \multicolumn{1}{c}{$g_7$}  
& \multicolumn{1}{c}{$g_8$}  
& \multicolumn{1}{c}{$\Lambda$}\\
\hline
9.834    & -122.9 & 6.189   & 4436.7  & 211.0   & -6647 & 1529     & 215.4             & -1666     &  29.81         & -63.20     & 0.8275\\\hline
8.731    &  -51.9 & 1.569   & 1009.9  & 411.8   & -2723 &  768     & 113.0             & -1162     & -10.31         & -85.13     & 0.6511\\\hline
\hline
  \multicolumn{1}{c}{$f_\pi$}
& \multicolumn{1}{c}{$f_K$}  
& \multicolumn{1}{c}{$\theta_{ps}$}
& \multicolumn{1}{c}{$\theta_{s}$}
& \multicolumn{1}{c}{$m_\pi$}
& \multicolumn{1}{c}{$m_K$}
& \multicolumn{1}{c}{$m_\eta$}
& \multicolumn{1}{c}{$m_{\eta^\prime}$}
& \multicolumn{1}{c}{$m_{a_0}$}
& \multicolumn{1}{c}{$m_{K^\star}$}
& \multicolumn{1}{c}{$m_\sigma$}
& \multicolumn{1}{c}{$m_{f_0}$}\\\hline
92       & 113    & -12           & 27.5         & 138     & 494   & 547      & 958               & 980       & 850           & 500        & 980\\\hline\\
\end{tabular*} 
\end{table*}

\begin{table*}
\caption{CEPs (position given as $\left\{\mu,T\right\}$) and the critical chemical potential, $\mu_{crit}$, at $T=0$ (sets a, b, c and d from \cite{Osipov:2013fka}; set e and f from Table \ref{parametersetD2}) for both transitions. All values are given in MeV. As one can see, comparing the results pertaining sets e and f, the position of the critical lines is almost unaffected by the use of a 3D hard cutoff in momentum space instead of Pauli-Villars regularization.}
\label{CEPS}
\center \begin{tabular*}{0.7\textwidth}{@{\extracolsep{\fill}}l|rr|rr@{}}
\hline
       \multicolumn{1}{c|}{Set}
     & \multicolumn{1}{c}{$\mathrm{CEP}_I$}
     & \multicolumn{1}{c|}{$\mu^I_{crit}|_{T=0}$}
     & \multicolumn{1}{c}{$\mathrm{CEP}_{II}$}
     & \multicolumn{1}{c}{$\left.\mu^{II}_{crit}\right|_{T=0}$}\\
\hline
a  & \{234.5, ~96.1\} & 332.0  & \{282.2 , 109.4\} & 410.6   \\\hline
b  & \{233.4, ~96.7\} & 332.0  & \{279.5 , 110.4\} & 410.0   \\\hline
c  & \{272.1, ~85.5\} & 343.6  & \{300.4 , 109.3\} & 423.6   \\\hline
d  & \{246.0, ~90.7\} & 332.8  & \{319.0 , 107.0\} & 434.8   \\\hline
e  & \{194.2, 104.7\} & 319.1  & \{307.8 , 106.2\} & 425.5   \\\hline
f  & \{216.2, 102.3\} & 331.5  & \{317.1 , 108.3\} & 442.0   \\\hline
\end{tabular*} 
\end{table*}

We trace back the existence of two critical lines mainly to the ordering $m_{K^\star}< m_{a_0}\approx m_{f_0}$  in the low lying scalar meson spectrum\footnote{In contradiction with this empirical observation the calculated scalar spectrum in the absence of the explicit $\chi_S$-breaking interactions usually displays the ordering $m_{a_0}<m_{K^\star}<m_{f_0}$ \cite{Dmitrasinovic:2000ei,Osipov:2006ns}.}. It has been shown analytically that the parameter $g_3$ plays a pivotal role in the assignment leading to the empirical masses of these mesons \cite{Osipov:2012kk, Osipov:2013fka}. The coupling $g_3$ is associated to a non OZI-violating term (see Subsection \ref{EffectL} and Equations \ref{Vst} and \ref{spc}) which counterbalances the flavour mixing 't Hooft interaction: we note, for instance, that by fitting the meson mass spectrum and weak decay constants to the values shown in Table \ref{parametersetD2}, but relaxing the constraint  for $m_{K^\star}$, a second critical endpoint (CEP) is still present  for $m_{K^\star}\simeq 953$ MeV (obtained by fixing $g_3=-1600~\mathrm{GeV^{-6}}$), but an increase  to values closer to $m_{a_0}=980$ MeV such as $m_{K^\star}\simeq 972$ MeV  ($g_3=-800~\mathrm{GeV^{-6}}$) leads to its disappearance as the additional first order transition changes into a crossover. It is the non-OZI violating character of $g_3$, not its explicit symmetry breaking, which matters the most in this consideration \footnote{If all flavor mixing terms were switched off the two sectors would decouple and two critical lines are obtained in the phase diagram.}.

That this transition is strongly correlated with the scalar sector should not be very surprising, being the scalars the carriers of the quantum numbers of the vacuum. Given the rather successful description of the scalar sector  (in addition to the pseudoscalars), ideal model conditions are thus provided to investigate the response of the vacuum subject to external conditions. In this way we expect to get a more complete picture of the role played by the explicit symmetry breaking interactions on the chiral transitions of the phase diagram. 

\begin{figure*}
\center
\subfigure[]{\label{grafMuTransition}\includegraphics[width=0.32\textwidth]{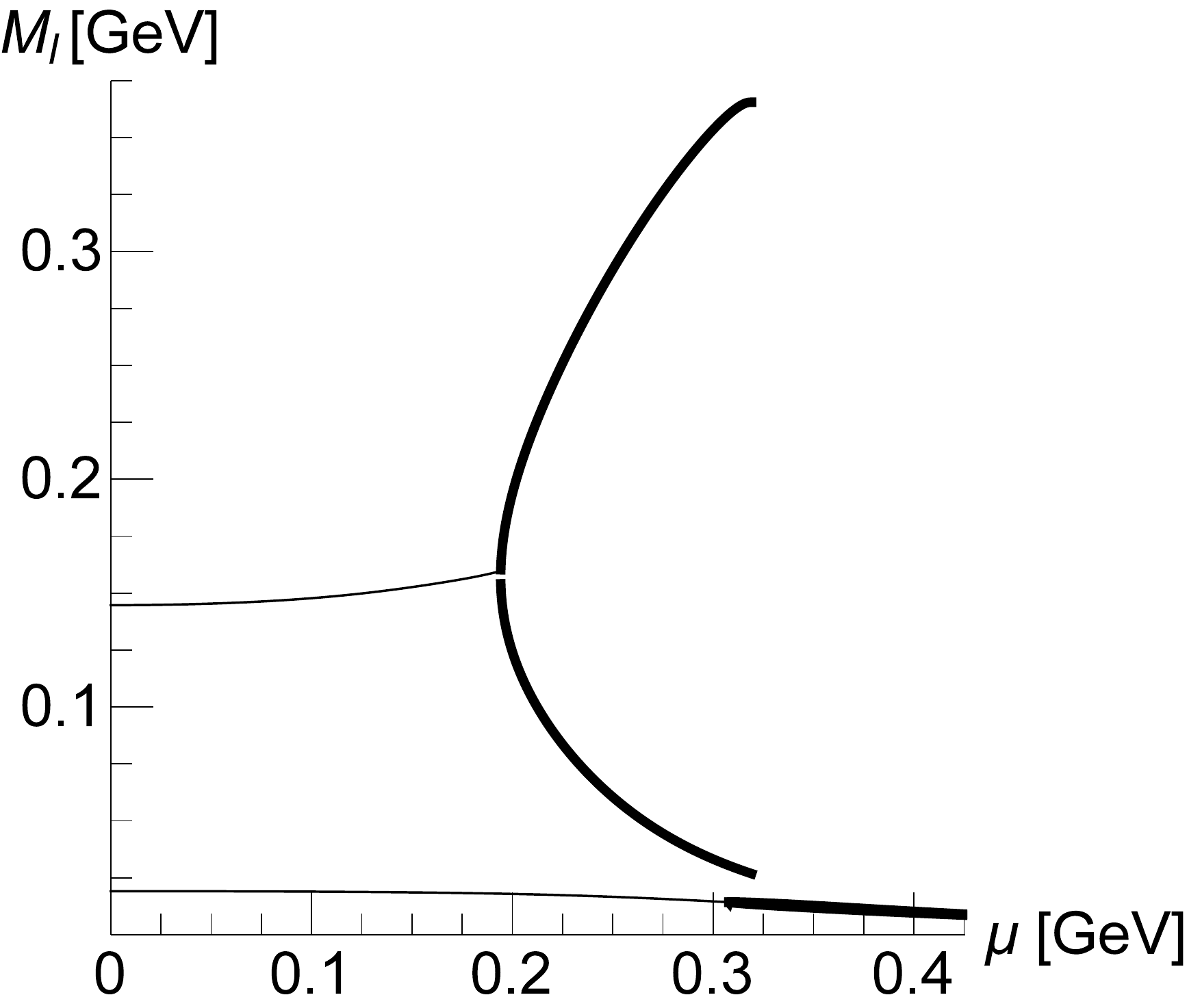}}
\subfigure[]{\label{grafMsTransition}\includegraphics[width=0.32\textwidth]{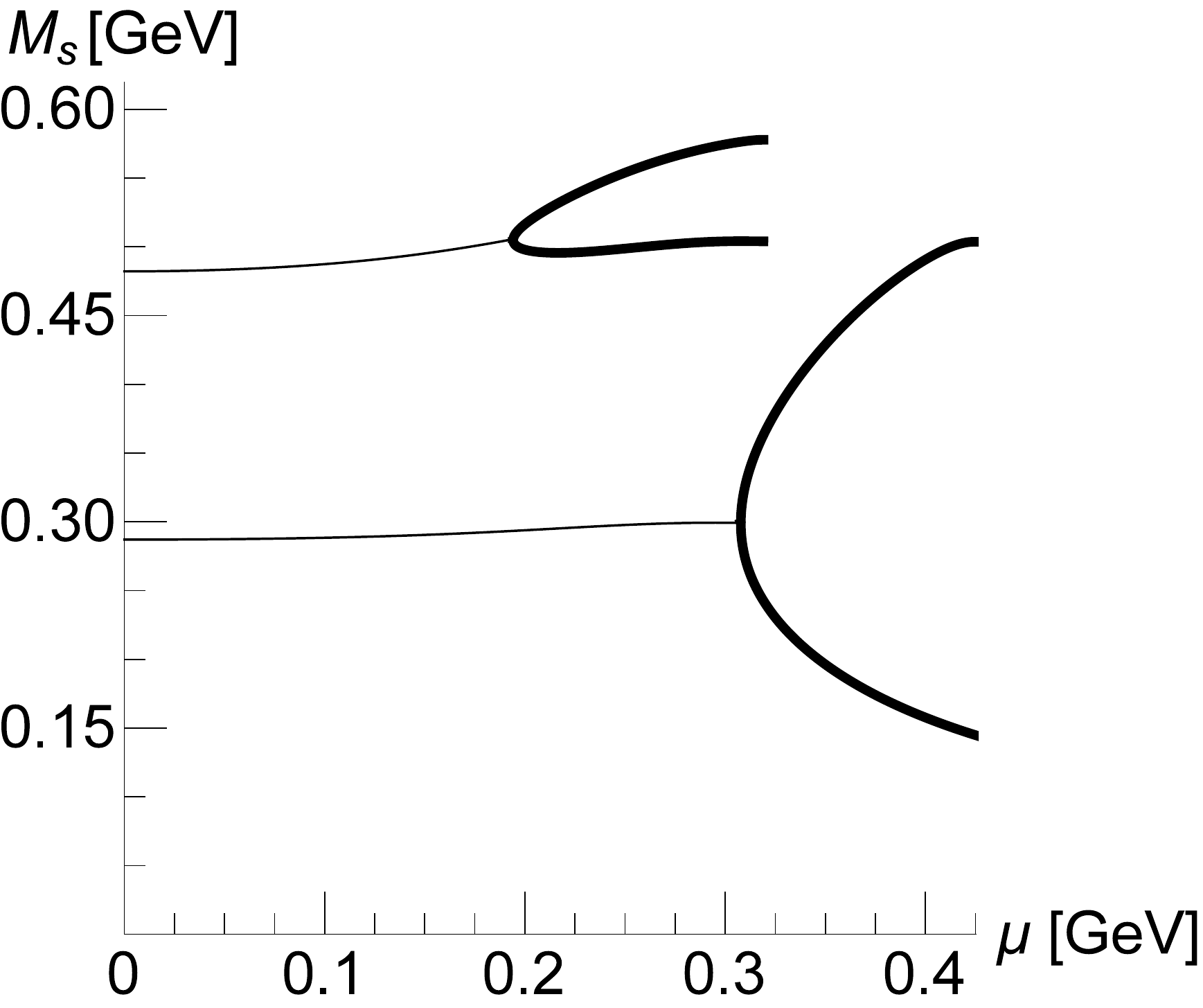}}
\caption{
Dynamical mass of the light quarks in \ref{grafMuTransition} and the strange in \ref{grafMsTransition} as a function of the chemical potential at the transitions obtained using the parameter set from Table \ref{parametersetD2} with $\mu_u = \mu_d = \mu_s = \mu$. Thin lines correspond to the value at the crossovers and thick lines to the masses in the first order transitions (at the first order transition a jump occurs from the upper to the lower branch). In each figure the lower set of lines corresponds to the outermost transition line in \ref{grafPDD2}. For the light quarks a much larger jump occurs at the transition corresponding to innermost line (the two thick lines corresponding to the outermost transition appear almost undistinguishable in the plot) whereas for the strange quarks the situation is reversed. The corresponding temperatures can be extracted from Fig. \ref{grafPDD2}.
}
\label{Imagens_TransitionMasses}
\end{figure*}

As the light and strange sectors are coupled by the OZI-violating interactions, all the quark masses are affected simultaneously by the two transitions (henceforth we will use the subscript $I$/$II$ when referring  to the one occurring at lower/higher $\mu$). Nevertheless a correspondence can be made between the transitions I/II and light/strange quarks as the chemical potential at which they occur at $T=0$ is relatively close to $M_l(0)$ and $M_s(\mu^I_{crit})$, respectively, and the jump is highly unequal in intensity (see Figs. \ref{Imagens_TransitionMasses}).

At $T=0$ the first order transitions occur at $\mu_I=0.319~\mathrm{GeV}$ and $\mu_{II}=0.426~\mathrm{GeV}$ connecting the phases with baryon number densities, $\rho=\left(\rho_u+\rho_d+\rho_s\right)/3$, given by: $\left\{ \rho_I^{-},\rho_I^{+}\right\}=\left\{0,1.664\right\}\rho_0$ and  $\left\{\rho_{II}^{-},\rho_{II}^{+}\right\}=\left\{3.984,5.643\right\}\rho_0$ (for the nuclear saturation density we use $\rho_0=0.17~\mathrm{fm}^{-3}$; the subscript $-/+$ refers to low/high density). 

For intermediate densities the system can be described by a mixture with a volume fraction, $0<\alpha<1$, occupied by the higher density phase. For densities in the range $\rho_I^{-}<\rho<\rho_I^{+}$, for instance, we expect a partial occupation of the volume by the zero pressure $\rho_I^+$ phase (with an energy per baryon of $E/A=958~\mathrm{MeV}$) in equilibrium with the vacuum. Pressure and chemical potential remain constant in this mixed phase regime. 

\subsection{\texorpdfstring{$\beta$}{Beta}-equilibrium case study}
\label{BetaEq}
To have a better understanding of the consequences of these two critical lines on the formation of SQM we now focus on charge neutral quark matter at $T=0$ subject to $\beta$-equilibrium, a case study of relevance for compact stars. Including charge neutrality this results in the conditions:
\begin{align}
\label{BetaEqCond}
\mu_u & =\mu-\frac{2}{3}\mu_e\nonumber\\
\mu_d &=\mu_s=\mu+\frac{1}{3}\mu_e\nonumber\\
    0 &=\frac{2}{3} \rho_u -\frac{1}{3}(\rho_d+\rho_s)-\rho_e
\end{align}
where the average quark chemical potential is given by $\mu=\frac{1}{3}\sum_i \mu_i$, $\mu_e$ is the electron chemical potential and  $\rho_i$ denotes number densities (which depend on the chemical potentials). The neutrino chemical potential is discarded as they are considered to escape the system. Muons need not be considered since the electron chemical potential remains below the muon mass in the present study. At $T=0$ the total energy density of the system is $\epsilon=\Omega +\sum_f \mu_f \rho_f$ (where $f=u,d,s,e$).
       
\begin{figure*}
\center
\subfigure[]{\label{grafDensT0000EqBeta}\includegraphics[width=0.32\textwidth]{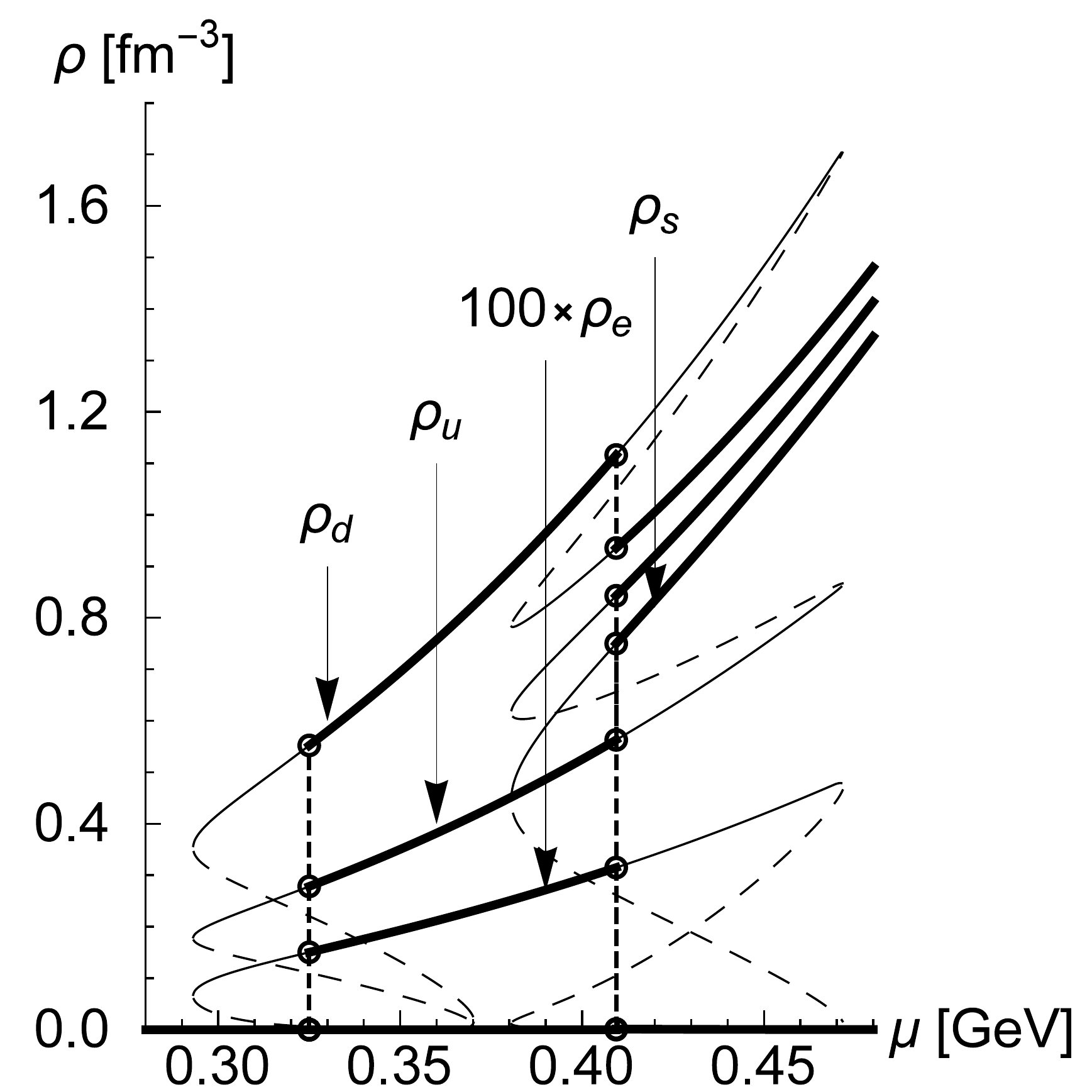}}
\subfigure[]{\label{grafDensAlphaFaseMistaEqbI}\includegraphics[width=0.32\textwidth]{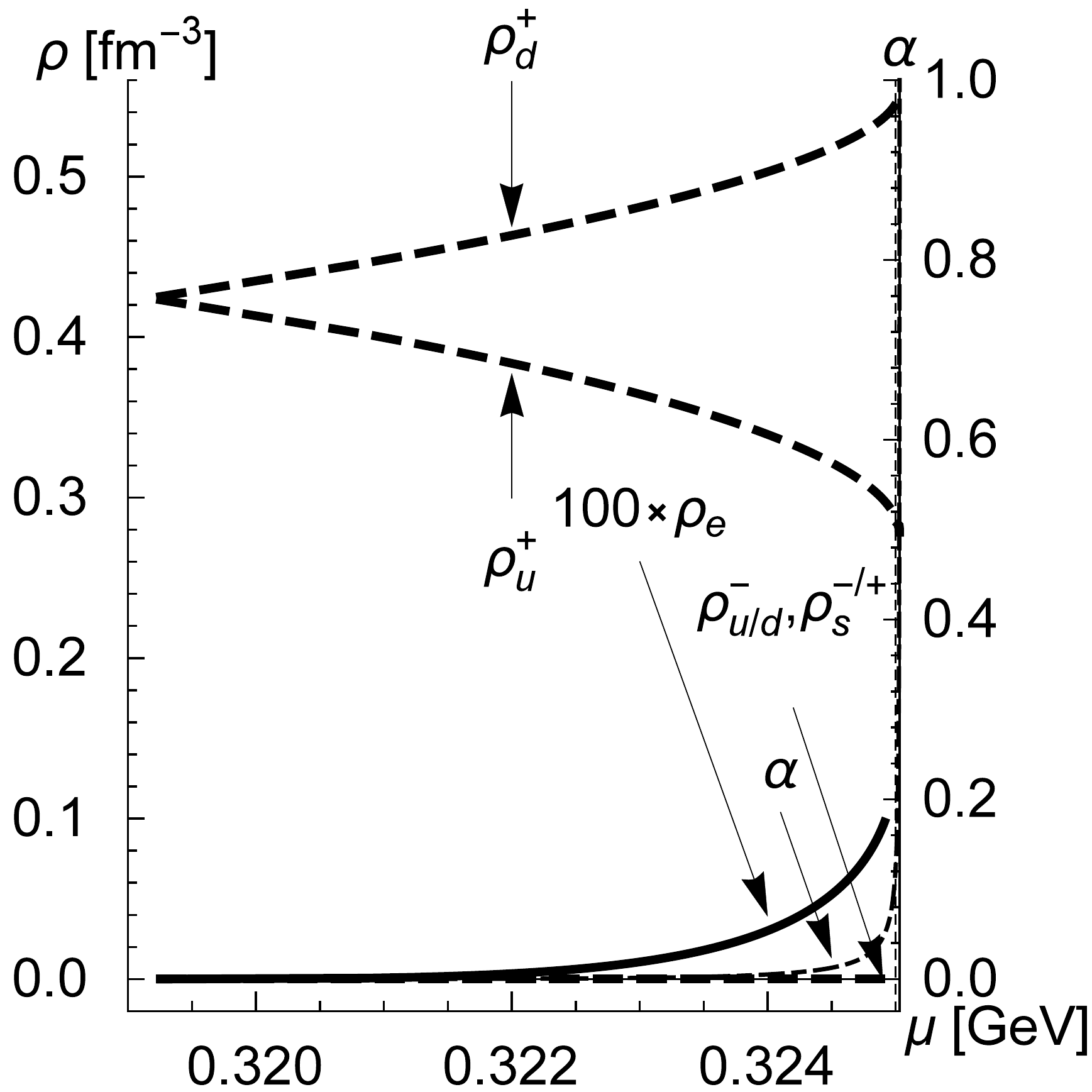}}
\subfigure[]{\label{grafDensAlphaFaseMistaEqbII}\includegraphics[width=0.32\textwidth]{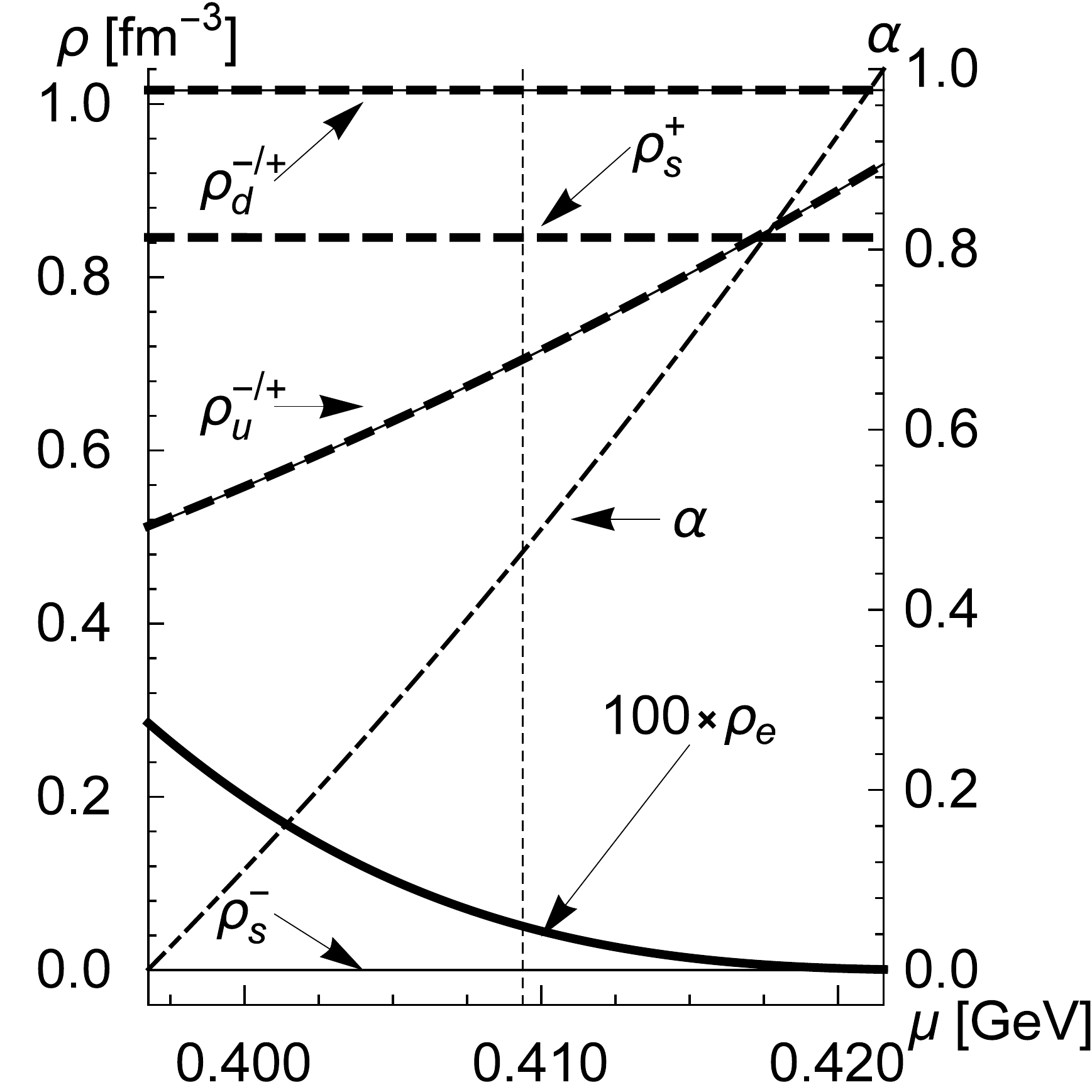}}
\caption{
Number densities for quarks ($u$, $d$ and $s$) and electrons (scaled by a factor of 100) as a function of the chemical potential, $\mu$, are shown in Fig. \ref{grafDensT0000EqBeta} for pure phases.  The mixed-phase constructions for both transitions using Gibbs conditions are described in Figs. \ref{grafDensAlphaFaseMistaEqbI} and \ref{grafDensAlphaFaseMistaEqbII} where besides the number densities of the species involved the volume fraction occupied by the emerging phases is displayed (the $-/+$ superscript refers to the lower/higher density phases). The thin vertical dashed lines refer to the critical chemical potentials of the first order transitions when only pure phases are considered. Note that on \ref{grafDensAlphaFaseMistaEqbI} both the last portion of the line refering to the denser phase volume fraction and the critical chemical potential are extremely close to the right-hand side axis and are as such barely visible.
}
\label{Imagens_EqBeta}
\end{figure*} 

As can be seen in Fig. \ref{grafDensT0000EqBeta} for chemical potential values lower than the dynamical masses the densities are trivially zero but as one increases $\mu$ the process of chiral restoration is initiated. After the lowest 1st order transition ($\mu_I=0.325~\mathrm{MeV}$) the Fermi sea gets populated by a finite density of up and down quarks, in chemical equilibrium with the electrons, while the density of strange quarks remains null until the highest 1st order transition ($\mu_{II}=0.409~\mathrm{MeV}$). For higher chemical potential the quark densities become comparable (a necessary condition to obtain SQM) while the electron density vanishes.

If we restrict ourselves to the consideration of pure phases these 1st order chiral transitions will give rise to jumps in the densities as a function of $\mu$: 
$\left\{\rho_I^{-}, \rho_I^{+}\right\} =\left\{0 ,1.628\right\}\rho_0$ and $\left\{\rho_{II}^{-},\rho_{II}^{+}\right\}=\left\{3.292,4.958\right\}\rho_0$.

The consideration of a mixed phase can however lower the total thermodynamic potential. This happens in the intervals of density: 
$0<\rho/\rho_0<1.629$ and $2.998<\rho/\rho_0<5.476$ (note that these intervals contain the ones corresponding to the jumps when considering pure phases). The main difference is that, with the inclusion of $\beta$-equilibrium, the chemical potential and pressure are no longer constant throughout the mixed phase \cite{Glendenning:1992vb} and the discontinuity of $\rho(\mu)$ disappears. One should note, nonetheless, that for each total density (or volume fraction) the pressure and chemical potentials of both phases must be equal for the system to be in mechanical and chemical equilibrium. Global charge  neutrality is assured by imposing (note that $\rho_e$ is constant throughout the volume):
\begin{align}
\label{GlobalCharge}
0=&\frac{2}{3}\left(\rho^-_u \left(1-\alpha\right)+\rho^+_u\alpha\right)\nonumber\\
  &-\frac{1}{3}((\rho^-_d+\rho^-_s)(1-\alpha)+(\rho^+_d+\rho^+_s)\alpha)-\rho_e.
\end{align}

This type of mixed phase construction where we consider two phases in chemical and mechanical equilibrium with global charge conservation (in this case charge neutrality) is a Gibbs construction. An alternative approach is to consider a Maxwell construction involving fractions of the volume occupied with the same electrically neutral phases that we obtain when considering only pure phases (with the same pressure and baryonic chemical potential but different individual electron and quarks chemical potentials).

A detailed treatment of a mixed-phase system involves the consideration of energy contributions coming from the interface between coexisting phases as well as electrostatic contributions. These however fall outside of the scope of the present study (for a more detailed discussion see for instance \cite{Bhattacharyya:2009fg} and references therein). In a Gibbs construction the interface is considered completely \emph{transparent} to the interchange of particles (only global charge neutrality is imposed) whereas in the Maxwell construction an interface between phases with different chemical potentials for the individual species must be considered (charge neutrality is imposed locally).

As in the present work we include no description of the interface between phases nor the electrostatic contribution, the Gibbs construction is energetically favourable, nevertheless for completeness we include both approaches in our study.

In Figs. \ref{grafDensAlphaFaseMistaEqbI} and \ref{grafDensAlphaFaseMistaEqbII} the chemical potential dependence of the number densities of quarks and electrons and of the volume fraction is shown for both transitions. The mixed phase transitions span very different ranges: for transition $I$ it spans $6~\mathrm{MeV}$ (with $\mu_I$ almost as its upper limit) and $25~\mathrm{MeV}$ for transition $II$ (roughly centered in $\mu_{II}$).  Furthermore, in the second transition the rise of $\alpha$ is approximately linear along the chemical potential interval whereas in the first transition the rise from $1/100$ to $1$ is achieved in the last $0.646~\mathrm{MeV}$ of the interval.

In the mixed phase corresponding to transition $I$ we see that in the limit of vanishing $\alpha$ the densities of up and down quarks are very close as one would expect from the $\beta$-equilibrium condition. The difference $\mu_d-\mu_u=\mu_e$ must go to the electron mass for the electron density to become increasingly small (see Eq. \ref{GlobalCharge}): for small $\alpha$ the charge imbalance from a small portion of quark matter is being compensated by a large volume of electron gas (in the low density phase we have $\rho^-_{I\,i}=0$ for $i=u,d,s$). As $\alpha$ approaches unity the density of down quarks almost doubles that of up quarks and we enter the regime seen in the intermediate interval of Fig. \ref{grafDensT0000EqBeta}. 

During the mixed phase associated with transition $II$ we see that the density of down quarks remains almost constant and approximately equal in both phases. The appearance of strangeness, only present in the denser phase, is compensated mainly by an increase in the density of up quarks which is also approximately equal in both phases. For $\mu<417~\mathrm{MeV}$ the denser phase becomes richer in strange quarks than in up quarks. Note that if we consider the total number density of species (taking into account both the high and low density phases) the ordering is always $\rho_d>\rho_u>\rho_s$.

\begin{figure*}
\center
\subfigure[]{\label{grafEpart}\includegraphics[width=0.32\textwidth]{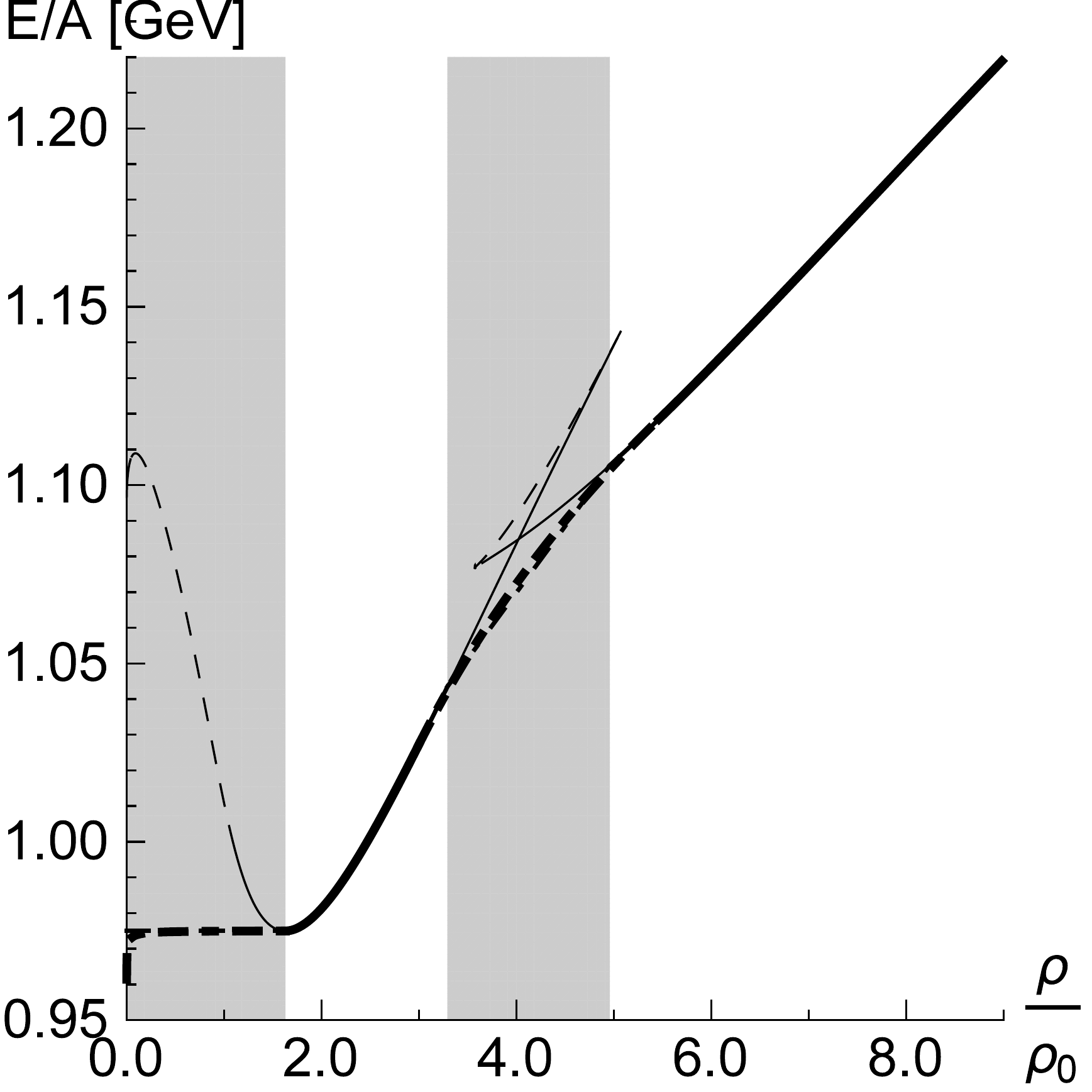}}
\subfigure[]{\label{grafEpartzooms}\includegraphics[width=0.32\textwidth]{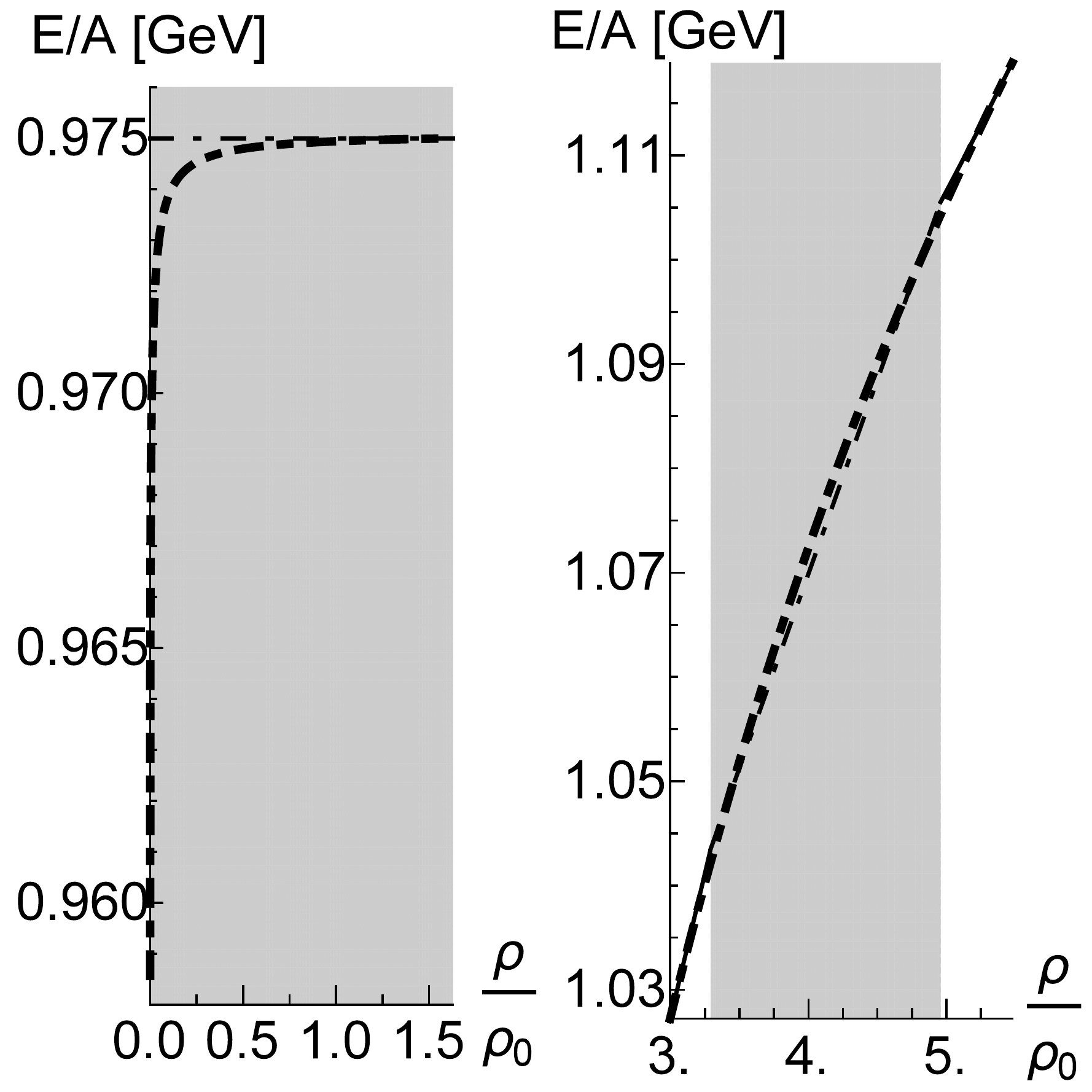}}
\subfigure[]{\label{grafpEner}\includegraphics[width=0.32\textwidth]{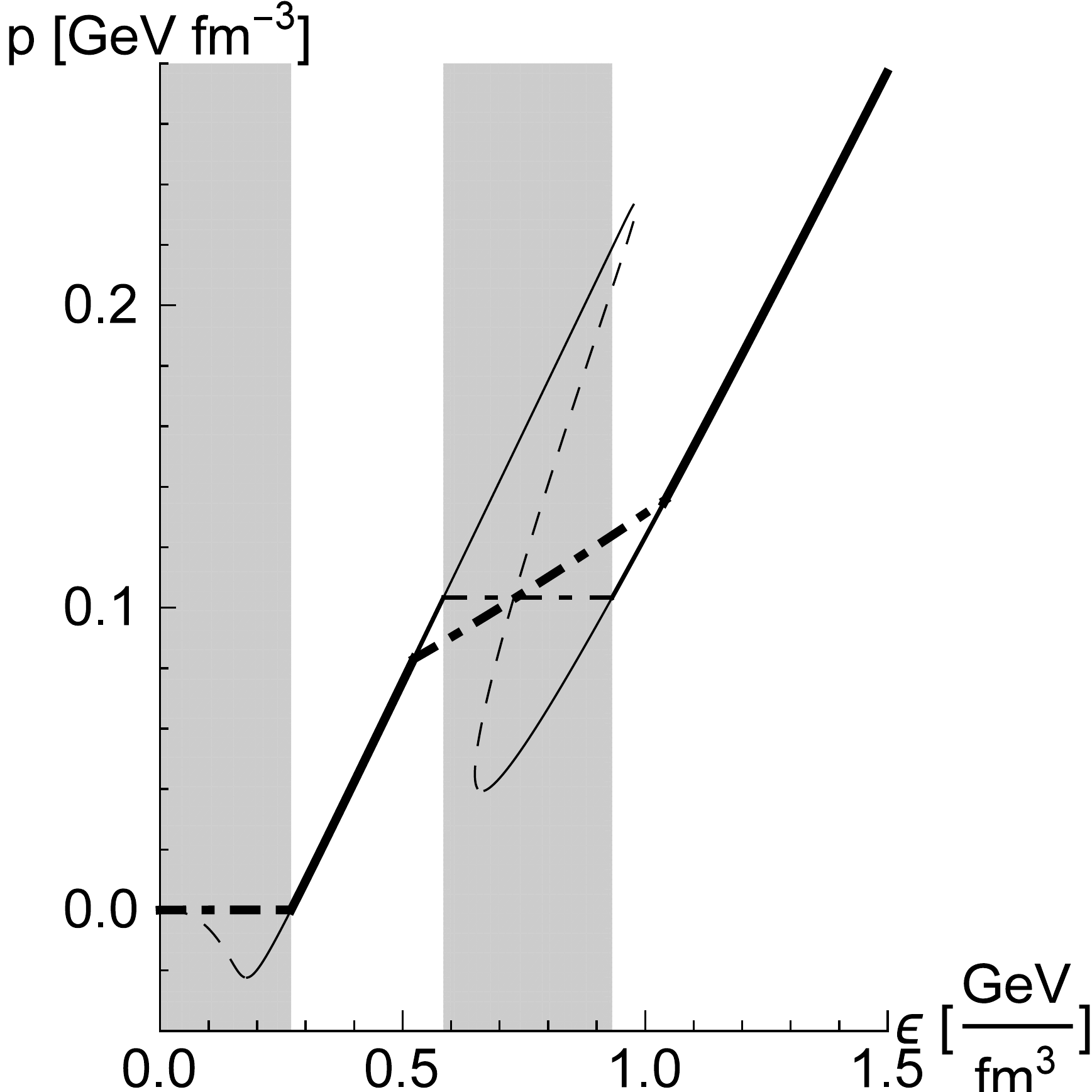}}
%\subfigure[]{\label{grafMassRadiusNJL8chiD2}\includegraphics[width=0.32\textwidth]{Imagens/grafMassRadiusNJL8chiD2}}
\caption{
In Fig. \ref{grafEpart} the energy per baryon number, $E/A=\epsilon/\rho$, as a function of baryon number density. Below the first order transition occurring at lower chemical potential there are two small regions of local minima solutions for $ 0<\rho/\rho_0<0.040$ and $ 1.051<\rho/\rho_0<1.628$ (thin full lines), the first of which is barely visible in the plot, connected by a zone of unstable solutions. The dot-dashed lines refer to the mixed phase constructions, a zoom of which can be seen in Fig. \ref{grafEpartzooms}. The thicker dot-dashed line refers to the Gibbs construction whereas the thinner refers to the Maxwell construction (which only exists in the grey areas). In the interval where they are both defined these constructions lead to lines which almost coincide (apart from the small density limit $\rho\lesssim 0.5\rho_0$).The state equation (pressure as a function of energy density) is shown in Fig. \ref{grafpEner}. The grey areas correspond to the jumps in density when we restrict the system to pure phases. The line notation (thin, thick and dashed) is the same as in previous figures.
% In Fig. \ref{grafMassRadiusNJL8chiD2} are shown the solutions of the TOV equations, the star masses (in units of solar masses) as functions of their radii. 
}
\label{Imagens_FasesMistas}
\end{figure*}

In  Figs. \ref{grafEpart} the energy per baryon $\frac{E}{A}$ as function of density is displayed. In the first transition mixed phase, as $\alpha$ goes to zero the energy per baryon changes from the $\beta$-equilibrium minimum value (when only considering pure phases the minimum is $E/A=975~\mathrm{MeV}$) to a value close to that without $\beta$-equilibrium ($E/A=959~\mathrm{MeV}$) as one expects from the fact that the densities of up and down quarks become very close (see discussion above). The second mixed phase connects smoothly the points: $\left\{\rho,E/A\right\}=\left\{2.998 \rho_0,\,1.027~\mathrm{GeV}\right\}$ and $\left\{\rho,E/A\right\}=\left\{5.476 \rho_0,\,1.119~\mathrm{GeV}\right\}$

At densities close to $\rho_0$ we have at most a metastable solution (at $1.051~\rho_0$ and $1.017~\rho_0$, with and without $\beta$-equilibrium, respectively). This inability of the model to describe bound nuclear matter at the nuclear saturation density is a well known property for the NJL model \cite{Buballa:1998pr} and can be related  to the lack of confinement. 

We call however attention to the fact that the onset of stable solutions, after the chiral transition of light quarks, occurs at densities closer to the nuclear density as compared to the values reported in \cite{Buballa:1998pr}, $\rho= 2.8\rho_0$ in the chiral limit and $\rho= 2.25\rho_0$ (with $m_u=m_d=5.5$ MeV, $m_s=140.7$ MeV of  \cite{Rehberg:1995kh}). Also at this point $E/A\simeq 975$ MeV, closer to the value of nuclear stability of iron, $930~\mathrm{MeV}$  as compared to the corresponding value in  \cite{Rehberg:1995kh}, $E/A \simeq 1100~\mathrm{MeV}$. 

A similar situation occurs at the 1st order transition leading to the onset of SQM solutions at values of $E/A$ much closer to nuclear matter stability energies as compared to the values obtained for the NJLH model in \cite{Buballa:1998pr}, where the smooth crossover of $M_s$ leads to values too large to support SQM. By extending the NJLH to include diquarks, a 1st order transition in $M_s$ also occurs \cite{Buballa:2003qv}, in connection to the 2SC/CFL transition. For the latter case the $E/A$ values have about the same magnitude in the region $\rho> 6\rho_0$ as the ones obtained in the present study. 

Despite the fact that certain pertinent aspects of compact stars, such as the effects of strong magnetic fields and rotational effects, have not been taken into account in the present study, one can use the obtained equation of state, EoS (pressure as function of the energy density, $p(\epsilon)$, see Fig. \ref{grafpEner}), in the integration of the Tolman–Oppenheimer–Volkoff equations \cite{Oppenheimer:1939ne, Tolman:1939jz} to obtain the mass and radius of a neutron star as a function of its central energy density (or pressure). The obtained maximum star mass falls short of the recently observed values of about two solar masses \cite{Demorest:2010bx,Antoniadis:2013pzd} but that should not be surprising due to these simplifications (furthermore the presence of more than one first order transition tends to soften the equation of state thus lowering the maximum mass). For this simplified scenario we obtain for the maximal star the following characteristics: a mass and radius of $M_{\mathrm{max}}=1.521~\mathrm{M}_\odot$, $R_{\mathrm{max}}=10.261~\mathrm{km}$  (the values reported in the literature using the NJLH derived EoS range from $1$ \cite{Bordbar:2012xf} to $1.45$ solar masses in \cite{Menezes:2009uc}) with a central pressure of $p_\mathrm{central}=0.128~\mathrm{GeV}\,\mathrm{fm}^{-3}$ (which corresponds to a baryonic density $\rho_\mathrm{central}= 5.144~\rho_0$). This central pressure leads to a core in the mixed phase with a SQM volume ocupation fraction of $\alpha=0.867$. The mixed phase core has a radius and mass of: $R_{\mathrm{core}}=4.029~\mathrm{km}$ $M_{\mathrm{core}}=0.166~\mathrm{M}_\odot$ (of which about one third are in the SQM phase, $M_{\mathrm{SQM}}=0.055~\mathrm{M}_\odot$).

As was previously mentioned the way the interface between phases in the mixed regime is considered is radically different in the Gibbs and Maxwell constructions. In the latter pressure is constant throughout the mixed phase regime and therefore that part of the EoS does not enter the integration of the TOV equation (a layer in the mixed phase would be squashed to vanishing thickness as no pressure gradient is present). Using the Maxwell construction the largest compact star is the one with a central pressure corresponding to transition II ($R_\mathrm{max}=10.412~\mathrm{km}$ and $M_\mathrm{max}=1.544~M_\odot$) and as such no stable stars with a SQM core exist in this case.

%Stars with a central pressure corresponding to a core with quark matter in phase $I$ lie along the thin solid line, whereas the ones in phase $II$ are along the instability branch shown as a dashed line. Stars with  a mixed core of SQM and light quark matter (Fig. \ref{grafDensAlphaFaseMistaEqbII}) are indicated by the bold solid line. 

Regarding the steepness of the EoS, a possible extension of the model, in line with the tower of relevant multi-quark interactions at NLO in $N_c$ counting, consists in the inclusion of the set of spin 1 interactions. It is known that four quark vector interactions stiffen the equation of state, however its onset is delayed, making hybrid stars with a quark core unstable; it was shown in a recent analysis of the SU(2) NJL model with four and eight quark spin 0 and spin 1 interactions that the higher order interactions allow to control stiffness without delaying the onset \cite{Benic:2014iaa}. The $SU(3)$ extension would provide information on the possibility of achieving the necessary stiffness with this model, while still allowing for SQM at the core of largest mass stars.

\section{Conclusions}
\label{Conclusions}
We conclude highlighting two main results:

By including all non-derivative terms relevant at the scale of chiral symmetry breaking (NJLH8m) which are of the same order in $N_c$ counting as the 't Hooft flavour determinant considered in the 3 flavour extension of the NJL model (NJLH), and using sets of parameters which describe the meson spectra of low-lying pseudoscalar and scalar meson nonets and the weak decay constants to great accuracy, we have obtained the chiral $T-\mu$ phase diagram of the model, which displays two critical lines and respectively two CEP. The second critical line is associated with the strange quark mass which undergoes a first order transition, in which it changes abruptly to values close to its current quark mass in a moderate chemical potential region, $\mu \simeq 410$ MeV at $T=0$, with strong consequences for SQM. 

When compared to previous studies based in the NJLH model the density at which SQM emerges using NJLH8m is lowered to: $\rho\simeq 4.0 \rho_0$ in the case of equal quark chemical potentials, $\rho\simeq 3.3 \rho_0$ in the case of pure phases in $\beta$-equilibrium and $\rho\simeq 3.0 \rho_0$ if we consider a Gibbs constructed mixed phase in $\beta$-equilibrium.

The energy per particle ratio of SQM is much lower than in the NJLH; values of similar magnitude as the ones obtained in NJLH8m are only reached if the NJLH model is enhanced with diquark interactions, as a consequence of a first order transition from the 2SC to the CFL phase. 

Our study can be refined to include the diquark interactions, although they cannot affect our central result, i.e. the model's ability to describe SQM. Diquarks will increase the number of critical points, opening the door for new phases in the region of relatively cold and dense quark matter. One may hope that the combined effect of explicit symmetry breaking interactions and diquarks enhances further the SQM formation, a subject certainly worth studying, but beyond the scope of the present work. 

The second main result is that the region for the minimum of quark matter stability ($\rho\simeq 1.7 \rho_0$, $E/A=958~\mathrm{ MeV}$ for the case with equal quark chemical potentials, $\rho\simeq 1.6 \rho_0$, $E/A=975~\mathrm{ MeV}$ for the case of pure phases in $\beta$-equilibrium and $\rho\simeq 1.7\rho_0$, $E/A=959~\mathrm{MeV}$ for the denser phase in a Gibbs constructed mixed phase in $\beta$-equilibrium) gets pushed much closer to the point of nuclear matter stability in comparison with other related model calculations.   

\section*{Acknowledgments}
Research supported by Centro de F\'{i}sica Computacional of the University of Coimbra,  by the Funda\c{c}\~ao para a Ci\^{e}ncia e Tecnologia grant SFRH/BPD/63070/2009, Or\c{c}amento de Estado and by the European Community-Research Infrastructure Integrating Activity Study of Strongly Interacting Matter (Grant Agreement 283286) under the Seventh Framework Programme of EU.

\bibliography{ELA_PhaseDiagram}
\end{document}